\documentclass[trackchanges]{aastex701}

\usepackage{hyperref}
\usepackage{amsmath}
\usepackage{enumerate}
\usepackage[utf8]{inputenc}
\usepackage{cleveref}
\usepackage{amssymb}
\usepackage[T1]{fontenc}
\usepackage{enumitem}
\usepackage{url}
\usepackage[misc]{ifsym}
\usepackage{multirow}
\usepackage{mhchem}
\usepackage{booktabs}
\usepackage{xeCJK}
\usepackage{subfigure}

\begin{document}

\title{First Lunar Farside SETI Observations for Periodic Signals with the Low-frequency Radio Spectrometer of Chang'E-4 Mission}

\author[0000-0002-1190-473X]{Jian-Kang Li (李健康)}
\affiliation{Institute for Frontiers in Astronomy and Astrophysics, Beijing Normal University, Beijing 102206, People's Republic of China}
 \affiliation{School of Physics and Astronomy, Beijing Normal University, Beijing 100875, People's Republic of China}
 \email{fakeemail1@google.com}  

\author[0000-0002-4683-5500]{Zhen-Zhao Tao (陶振钊)}
 \affiliation{Institute for Astronomical Science, Dezhou University, Dezhou 253023, People's Republic of China}
 \affiliation{College of Computer and Information, Dezhou University, Dezhou 253023, People's Republic of China}
 \email{fakeemail1@google.com}  

 \author[0000-0002-8719-3137]{Bo-Lun Huang (黄博伦)}
\affiliation{Institute for Frontiers in Astronomy and Astrophysics, Beijing Normal University, Beijing 102206, People's Republic of China}
 \affiliation{School of Physics and Astronomy, Beijing Normal University, Beijing 100875, People's Republic of China}
 \email{fakeemail1@google.com}  

 \author[0000-0003-0167-9345]{Kang-Jiao (焦康)}
\affiliation{Institute for Astrophysics, Zhengzhou University, Zhengzhou 450001, People's Republic of China}
 \email{fakeemail1@google.com}  

\author[0000-0002-3363-9965]{Tong-Jie Zhang (张同杰) \href{mailto:tjzhang@bnu.edu.cn}{\textrm{\Letter}}}
\affiliation{Institute for Frontiers in Astronomy and Astrophysics, Beijing Normal University, Beijing 102206, People's Republic of China}
\affiliation{School of Physics and Astronomy, Beijing Normal University, Beijing 100875, People's Republic of China}
\email[show]{tjzhang@bnu.edu.cn}

\author{Ming-Yuan Wang (王明远)}
\affiliation{National Astronomical Observatories, Chinese Academy of Sciences, Beijing 100101, People's Republic of China}
 \email{fakeemail1@google.com} 

\author{Jin-Song Ping (平劲松)}
\affiliation{National Astronomical Observatories, Chinese Academy of Sciences, Beijing 100101, People's Republic of China}
 \email{fakeemail1@google.com} 

\author{Dan Werthimer}
\affiliation{Space Sciences Laboratory, University of California, Berkeley, CA 94720, USA}
\email{fakeemail1@google.com} 

\author{Vishal Gajjar}
\affiliation{SETI Institute, 339 Bernardo Ave, Suite 200, Mountain View, CA 94043, USA}
\email{fakeemail1@google.com}

\begin{abstract}
Chang'E-4 (CE4), the first mission to soft-land on the lunar farside, provides a unique opportunity for astronomical observations from an environment shielded from terrestrial radio interference, and thus serves as pathfinder for lunar farside radio search for extraterrestrial intelligence (SETI) studies. We present a search for periodic technosignatures using low-frequency radio observations from the CE-4  mission, the first radio SETI study based on data from on the observation in lunar farside. We analyze the CE4 dynamic spectra with a component-level framework that combines principal component analysis (PCA), cross-antenna basis alignment, as well as temporal periodicity and frequency comb structure diagnostics. No final periodic candidate signal is found after the selection procedure, and we therefore find no evidence in the present CE4 sample for a credible periodic artificial signal. This study serves as a pathfinder and provides a practical framework for lunar radio SETI analysis. As more future lunar missions begin to incorporate radio instrumentation, lunar farside may become a promising site for expanding radio SETI research.

\end{abstract}



\section{Introduction} \label{sec:Introduction}

The search for extraterrestrial intelligence (SETI) seeks observational evidence of technological activity beyond Earth and addresses one of the most fundamental questions in science: whether intelligent life exists elsewhere in the Universe. Radio SETI has long been regarded as one of the most promising aspects, because engineered transmissions could be transmitted efficiently with low energy requirements and minimal propagation losses \citep{1959Natur.184..844C,2001ARA&A..39..511T}.

Radio SETI searches have traditionally concentrated on narrowband and frequency-drifting signals \citep[e.g.][]{2013ApJ...767...94S,2017ApJ...849..104E,2020AJ....159...86P,2020AJ....160...29S,2021AJ....161...55M,2021AJ....161..286T,2022AJ....164..160T}, which are often considered energetically efficient and comparatively distinguishable from known natural astrophysical emission. 
Yet there is no need to restrict technosignature to this category.
Signals with repeatable or periodic modulation patterns in time also deserve  being included as an interest type for SETI study, because they represent a plausible extension of conventional form of structured artificial transmission. Previous searches have also considered periodic or pulsed forms of technosignatures \citep{2018ApJ...869...66H,2023AJ....165..255S,2026AJ....171...78L}, suggesting that the relevant observational space extends beyond conventional narrowband drifting signals.

One of the central challenges in SETI is the identification and rejection of radio frequency interference (RFI) in real observations. Terrestrial radio observations are strongly affected by anthropogenic transmissions generated by a wide range of electronic systems, making external RFI a persistent limitation for SETI experiments. In this respect, the Chang'E-4 (CE-4) mission provides a distinctive opportunity for technosignature search. Operating on the lunar farside, it benefits from a much quieter electromagnetic environment, where the Moon strongly suppresses direct RFI contamination originating from Earth \citep{2012ExA....33..529M,2020arXiv200912689M}. As a result, the observation data are expected to be largely free from external RFI, substantially reducing one of the major difficulties that usually accompanies SETI observations on Earth.

In this work, we utilize principal component analysis (PCA) approach to search for periodic signals with finite channel width in the CE-4 data. This PCA-based method is used to separate dynamic spectral data into a set of independent component bases, after that we characterize the temporal periodicity of the resulting components, and compare their behavior across different antennas by cross-antenna basis alignment to assess whether the components are shared among antennas, hence to identify whether periodic components are consistent with cross-antenna interference-like behavior. We introduce the observation details of the low-frequency radio spectrometer (LFRS) in the CE-4 mission in Section \ref{sec:Observations}. The methods we used for data analysis in this work are shown in Section \ref{sec:DataAnalysis}, and the corresponding results are presented in Section \ref{sec:Results}. Section \ref{sec:Discussion} discusses the implications of our results and Section \ref{sec:Conclusion} lists the conclusions of this work.

\section{Observations} \label{sec:Observations}

\begin{figure}[htpb]
  \centering
  \includegraphics[width=0.75\textwidth]{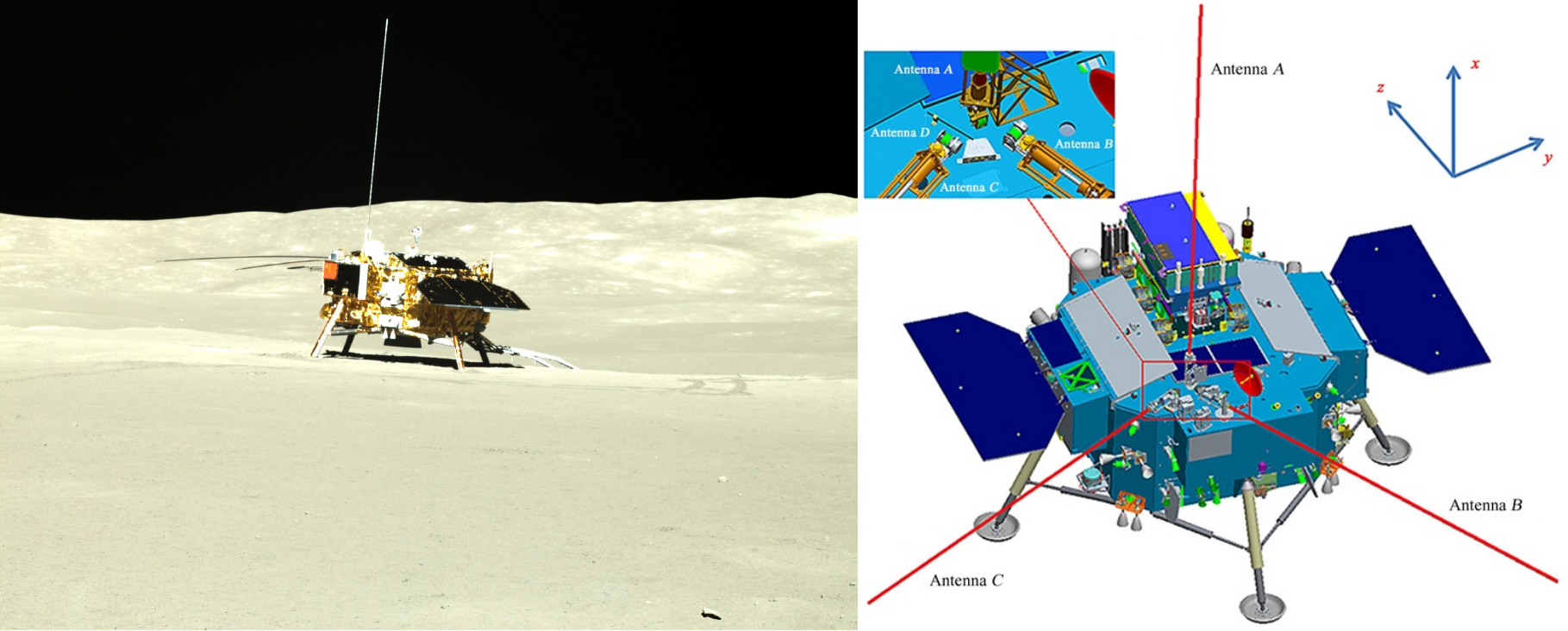}
  \caption{Real working environment and schematic diagram of CE-4 lander. The photo of CE-4 lander in the left panel was taken by the panoramic camera of CE-4 cruiser (Yutu-2). The antenna distribution of LFRS is illustrated in the right panel.}
  \label{fig:CE4_GRAS_PCAML}
\end{figure}

CE-4 mission landed on the eastern floor of Von K\'{a}rm\'{a}n crater within South Pole-Aitken basin \citep{2019NatGe..12..222W}. The LFRS is equipped on the top of CE-4 lander, consisting of three orthogonal 5-meter long dipole antennas denoted as A, B and C, as well as a 20-cm short antenna D parallel to the lander wall (See Figure \ref{fig:CE4_GRAS_PCAML}). The antenna D is mainly designed for background noise mitigation, as well as data calibration for other three dipole antennas. The LFRS works within frequencies of 100 kHz-2 MHz (low band) and 1-40 MHz (high band), with receiver sensitivity $\leqslant 10 \mathrm{nV/\sqrt{Hz}}$. The frequency resolutions are $\leqslant 10$ kHz for low band and $\leqslant 200$ kHz for high band \citep{2021RAA....21..116Z}. 

The LFRS started its observation on Jan 5 2019, and collected data for several years. Because these dipole antennas are mounted with fixed orientations on the lander, the sky coverage is determined passively by the fixed antenna geometry together with the lunar motion rather than being actively pointed toward a selected target. As a result, the three antennas provide complementary views of the radio environment, which are later used in our comparison of signal components across antennas.

In this paper we use the 2C-level science data products\footnote{The LFRS 2C data of CE4 mission can be downloaded from \url{https://moon.bao.ac.cn/mall/moonDATA}.} stored as binary files in PSD format, while the corresponding 2CL product label files are provided as XML-based ASCII files containing the observational information. Each 2C file consists of a sequence of science records, with each record containing a 2048-channel power spectral density measurement together with auxiliary metadata fields such as instrument status and observing geometry. In this work, the spectral arrays are extracted from these records and grouped by antenna, so that separate time-frequency dynamic spectra can be constructed for the three fixed antennas.

In many SETI observations, candidate signals are evaluated through dedicated on-source and off-source measurements, with the off-source data serving as a reference for identifying RFI. However, because the antennas of CE4 LFRS are fixed on the lander, the observations are closer to a passive drift-scan configuration, in which the sky coverage is set by the observing geometry rather than by active switching between target and reference pointings. We therefore compare the components derived from the dynamic spectra of three antennas, and search for shared behavior among them, using such common components as one indication of instrumental or RFI-like contributions.

\section{Data Analysis} \label{sec:DataAnalysis}

\subsection{Principal Component Analysis}\label{subsec:PCA}
Each observation can be represented as a two-dimensional dynamic spectrum, which can be regarded as a matrix whose axes correspond to time and frequency. PCA provides a natural way to transform and decompose such a matrix into a set of dominant orthogonal modes. For the search for time-periodic signals with finite channel width considered in this work, PCA can separate the major patterns in the data into temporal coefficients and corresponding spectral structures, allowing periodic behavior in time and finite-bandwidth structure in frequency to be examined consistently.

For a dynamic spectrum $X \in \mathbb{R}^{N_t \times N_\nu}$ with $N_t$ being the number of time samples and $N_\nu$ he number of frequency channels, it can be decomposed into
\begin{equation}
  X = U\,\Sigma\, V^T,
\end{equation}
where the columns of $U\in\mathbb{R}^{N_t\times r}$ describe the temporal behavior of the modes, the rows of $V\in\mathbb{R}^{N_\nu\times r}$ give the corresponding spectral structures, $\Sigma=\mathrm{diag}(\sigma_1,\ldots,\sigma_r)$ is the rectangular diagonal matrix containing their singular values and $r\le \min(N_t,N_\nu)$ is the rank of the matrix. In this representation, the temporal coefficients can be examined for periodicity, while the corresponding components describe its frequency-domain structure.

We apply PCA separately to the data from each antenna. The number of components for reconstructing the dynamic spectrum is determined from the cumulative explained variance ratio (CEVR) by requiring that the retained modes account for at least 99\% of the total variance. In practice, we found that the CEVR curve generally becomes much flatter near this level as the number of components increases. In our search, the signals we are interested in consist of time-periodic features with finite channel width. Such structures can therefore be separated into individual components after decomposition, where their temporal periodicity and spectral extent can be examined separately. Candidate periodic signals may also become more evident in individual components than in the full dynamic spectrum.

\subsection{Time and Frequency Domain Characterization} \label{subsec:period_and_freq_domain}

After the PCA decomposition, each retained component is characterized through its temporal and spectral behavior. The temporal part of a component is described by the corresponding mode in $U\Sigma$, while the frequency part is described by the associated column of $V$. To search for periodic structure in time, we compute a periodogram for the temporal part of each component and identify the dominant peak period $p_0$ in the periodogram. Components with clear peaks in the temporal spectrum are treated as potentially relevant to the periodic signals we search. In this work, The searched period range is bounded as $[6t_{\rm samp}, \tau_{\rm obs}/3]$, where $t_{\rm samp}$ is the sampling interval and $\tau_{\rm obs}$ is the total observation duration. The lower bound is chosen to ensure that each cycle is sampled by several time bins, while the upper bound is chosen to require that at least a few cycles are contained within a single observation. On the other hand, this upper bound should be regarded as an algorithmic search limit rather than a physically unconstrained signal timescale. Because the CE4 observations are effectively drift-scan rather than pointed tracking observations, an extraterrestrial signal from a fixed sky direction would remain within the effective beam response only for a limited dwell time. Periodicities much longer than this timescale are therefore difficult to interpret as arising from a sky-fixed ETI transmitter.

We also characterize the frequency-domain behavior of each component by analyzing the associated frequency vector $V$ using both a Fourier transform (FT) and an autocorrelation function (ACF). Both diagnostics are used to test whether the frequency component exhibits comb-like structure in frequency, which can be one of the possible instrumental RFI in the observation \citep{2023RaSc...5807595L}.

\begin{figure}[htpb]
  \centering
  \includegraphics[width=0.75\textwidth]{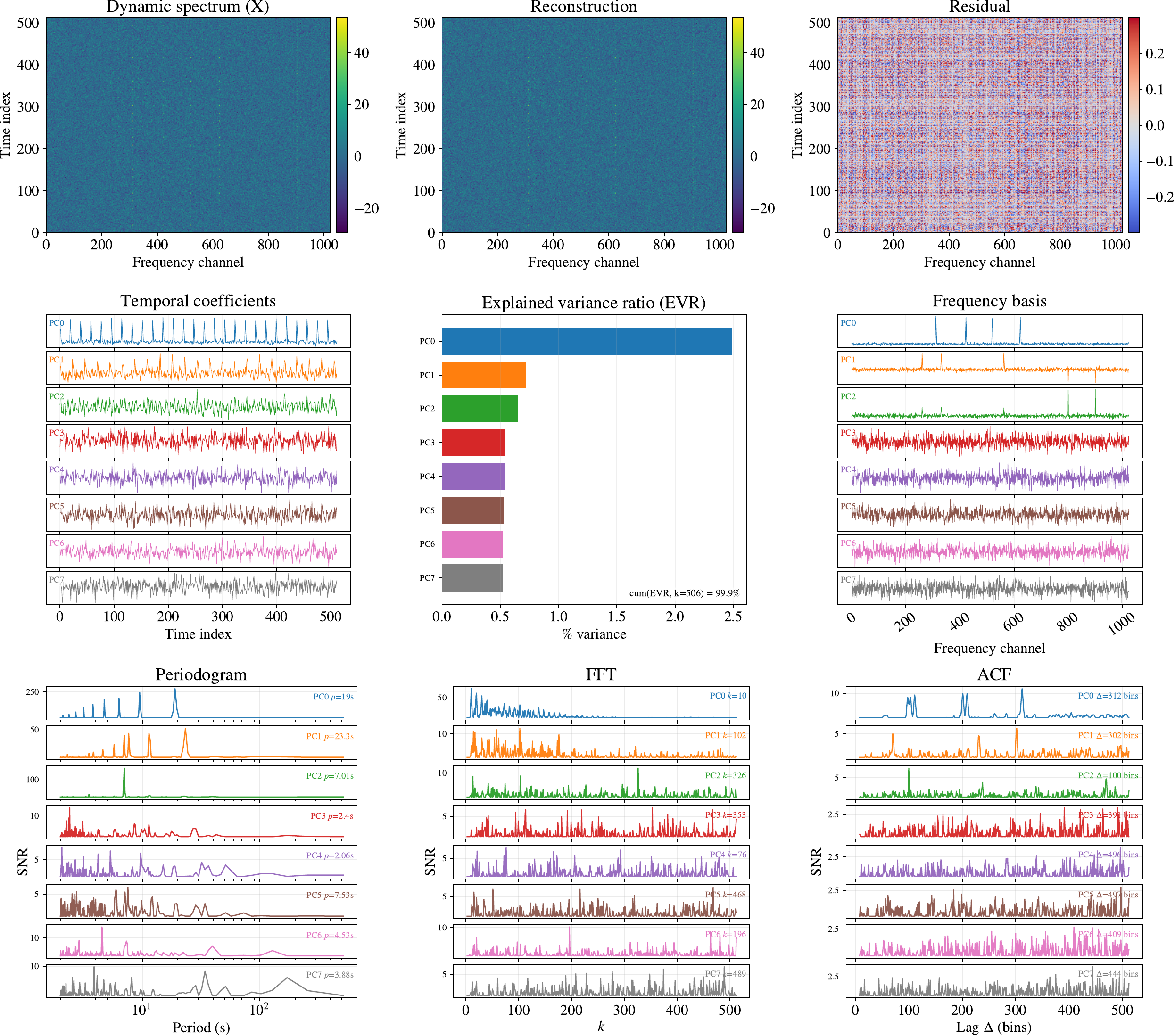}
  \caption{Example PCA diagnostic plot for simulated channel-width periodic signals. In this simulated example, we inject three periodic signals in some frequency channels: (i) $P=23$ s in three random frequency channels; (ii) $P=19$ s in four frequency channels with equal space (comb structure); (iii) $P=7$ s in two random frequency channels. Top row shows the original dynamic spectrum (left), the reconstructed spectrum by retained components (middle), and the residual (right). Middle row shows the temporal coefficient (left), explained variance ratio (middle), and the frequency basis vector of each component. Bottom row shows the periodogram of temporal coefficient (left), FT spectrum (middle) and ACF profile (right) of frequency basis for each component.}
  \label{fig:example_periodic_PCA}
\end{figure}

For the above temporal and frequency diagnostics, the prominence of a feature is quantified through a robust signal-to-noise ratio (S/N) defined relative to the background of the corresponding statistic. Specifically, for a quantity $x$, we define
\begin{equation}
{\rm S/N} = \frac{x - {\rm median}(x)}{1.4826{\rm MAD}(x)},
\end{equation}
where the median and median absolute deviation (MAD) are used to provide a robust estimate of the background level. This definition is applied consistently to the temporal periodogram, as well as to the FT and ACF analysis of the frequency components, so that significant periodic or comb-like features can be identified in a uniform way. And example for the simulated periodic signal search with PCA temporal and frequency diagnostics are illustrated in Figure \ref{fig:example_periodic_PCA}. We show only the top principal components in the PCA summary figures, since these leading components already account for the main variance contribution to the two-dimensional dynamic spectrum, whereas the higher-order components, although still included in the subsequent analysis and quantitative diagnostics, contribute only weakly to the matrix reconstruction.

\subsection{Cross-antenna Basis Alignment}
In the absence of reference off-source observation, an important part of the present analysis is to determine whether similar components are shared among the three antennas. Such shared components are of particular interest because they may reflect common interference-like structure, rather than signals associated with a specific observing direction. 

To assess whether a component identified in one antenna has a genuinely corresponding counterpart in another antenna, we therefore perform a cross-antenna basis alignment for quantitative assess of temporal and frequency bases across antennas through similarity measures for time $S_t$ and for frequency $S_\nu$. These similarities can be converted into test statistics for temporal and frequency alignment, which are then combined into a joint statistic used to assign p-values to matched pairs. The detailed derivation for basis alignment is given in Appendix \ref{Appendix:basis_alignment}.

\subsection{Selection of Candidate Non-RFI Periodic Signals}

After PCA decomposition, each retained component is characterized through its temporal periodogram, spectral FT spectrum and ACF profile, along with cross-antenna behavior is evaluated through basis alignment. In addition, we perform harmonic-prominence tests on the temporal periodicity and comb-prominence tests on its frequency structure.

In the temporal domain, if the dominant period of a component is $p_0$, then harmonically related peaks are expected at
\begin{equation}
p_n = \frac{p_0}{n}, \qquad n=1,2,\dots,H .
\end{equation}
For a non-sinusoidal periodic structure with finite duty cycle $D$, the harmonic strengths are expected to follow an ordered envelope,
\begin{equation}
S_n \propto \left[\frac{\sin(\pi n D)}{\pi n}\right]^2 ,
\end{equation}
where $S_n$ denotes the peak strength at the $n$th harmonic location. We therefore test not only whether a dominant periodic peak is present, but also whether the associated harmonic peaks are consistent with the expected relations in both position and relative strength.

In the frequency domain, if a component contains comb-like structure with characteristic spacing $\Delta$ along the frequency axis, then the corresponding peaks in the FT spectrum are expected to satisfy
\begin{equation}
k_m = m k_0, \qquad m=1,2,\dots,M ,
\end{equation}
where $k$ labels the characteristic repetition scale of the frequency structure in the FT spectrum. For a comb spacing of $\Delta$ frequency bins, the dominant mode satisfies approximately
\begin{equation}
k_0 \approx \frac{N_\nu}{\Delta},
\end{equation}
with $N_\nu$ being the number of frequency channels. In the ACF profile, the same repeated structure is expected to produce peaks at
\begin{equation}
L_m = m \Delta, \qquad m=1,2,\dots,M ,
\end{equation}
where $L_m$ is the lag in frequency bins. These relations are used to test whether the frequency structure of a component is consistent with repeated or approximately equally spaced comb-like behavior.

\begin{figure}[htpb]
  \centering
  \includegraphics[width=0.75\textwidth]{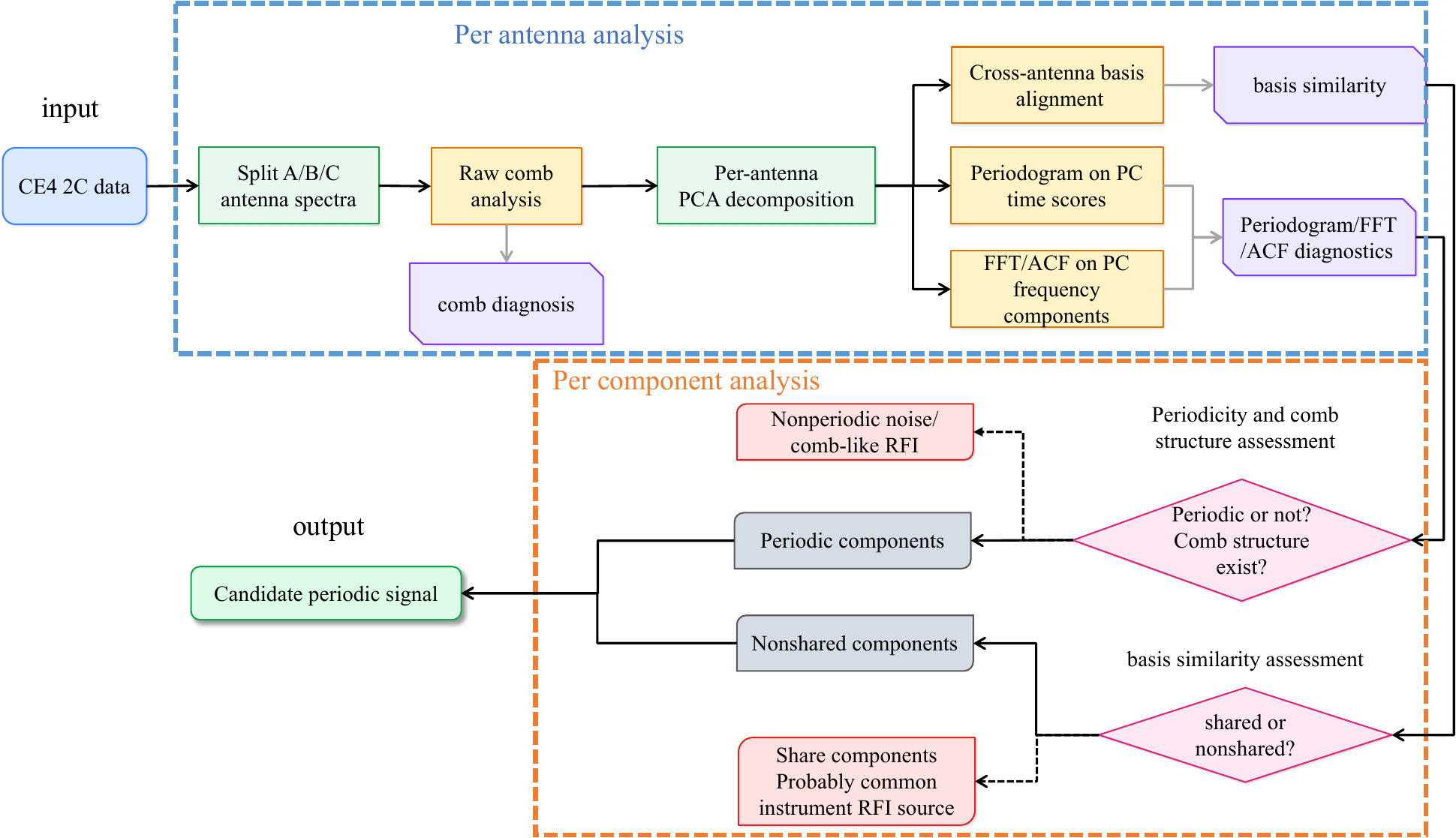}
  \caption{Schematic overview of the data-processing workflow in this work. After splitting the 2C data into three antenna spectra, we first conduct a general comb diagnostics for each antenna, then perform per-antenna PCA decomposition, basis alignment, and periodogram/FT/ACF analysis. The component-level results are finally combined to distinguish non-shared periodic signals from shared or comb-like RFI.}
  \label{fig:CE4_flowchart}
\end{figure}

The final non-shared periodic candidates are selected in a post-processing step by combining the temporal harmonic test, the cross-antenna basis-alignment result, the frequency-domain comb test, and the adopted SNR cuts. The full candidate-selection criteria are summarized in Table~\ref{tab:selection_parameters}, and the general workflow for the data analysis in this work is illustrated in Figure \ref{fig:CE4_flowchart}.

\begin{deluxetable*}{ll}[htpb]
\tablecaption{Adopted thresholds and parameters used in the selection of non-shared periodic candidates.\label{tab:selection_parameters}}
\tablehead{
\colhead{Selection criterion} & \colhead{Adopted value}
}
\startdata
Periodogram peak prominence threshold & $10\sigma$  prominence \\
Periodogram minimum peak distance & 8 bins \\
FT peak prominence threshold & $3\sigma$  prominence \\
FT minimum peak distance & 3 bins \\
FT matching window & 10 bins \\
ACF peak prominence threshold & $3\sigma$  prominence \\
ACF minimum peak distance & 3 bins \\
ACF matching window & 10 bins \\
Cross-antenna alignment significance threshold & $\alpha = 0.01$ \\
\enddata
\tablecomments{Here the prominence is defined relative to the median background level and the MAD-based scatter of the corresponding diagnostic.}
\end{deluxetable*}

\section{Results} \label{sec:Results}

\begin{figure}[htpb]
  \centering
  \includegraphics[width=0.85\textwidth]{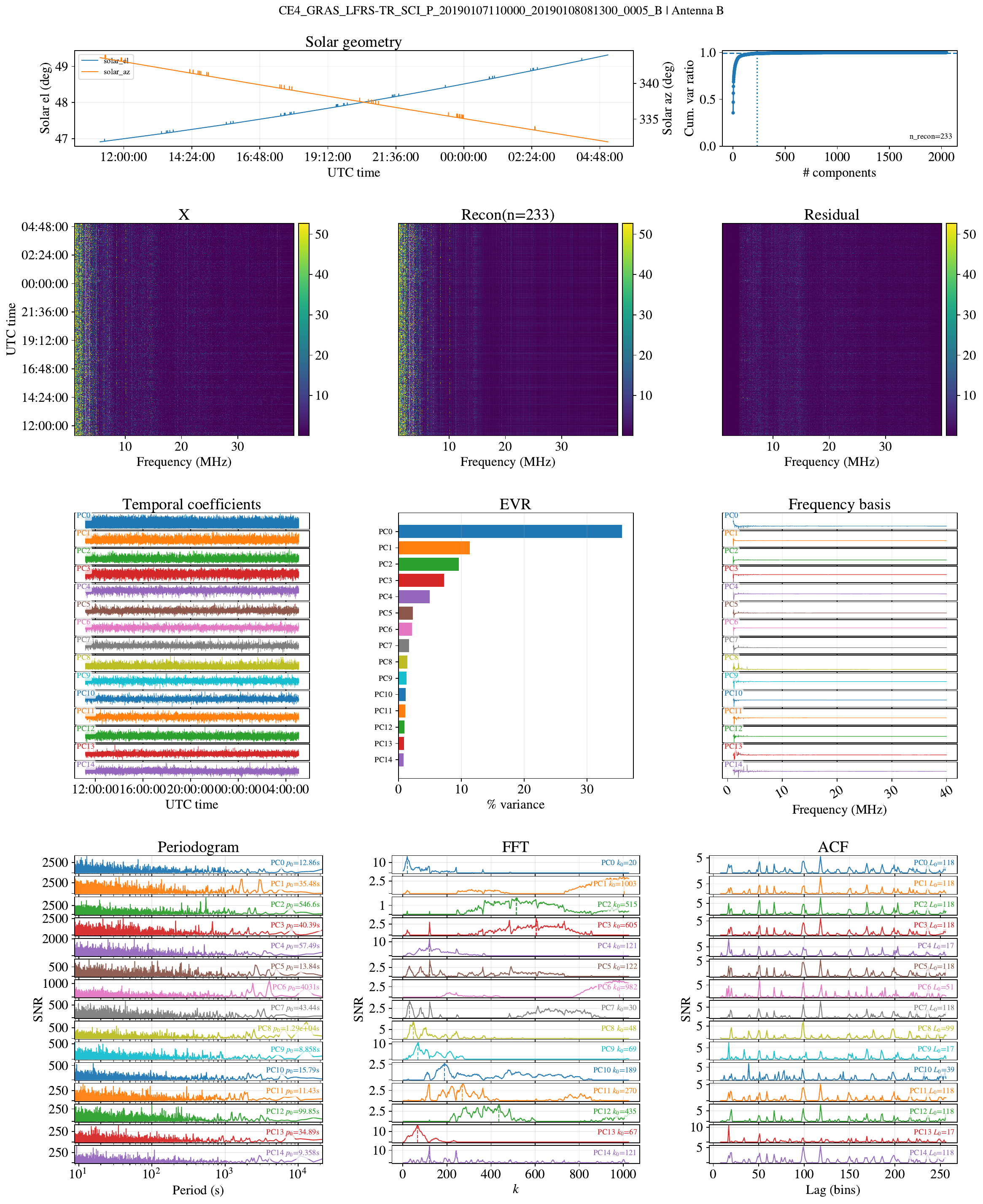}
  \caption{Example PCA diagnostic summary plot for real observation data. The general layout for the plot is similar to Figure \ref{fig:example_periodic_PCA}. At the top we also add the solar celestial coordinates records as well as the CEVR profile for the observation data.}
  \label{fig:example_PCA_summary}
\end{figure}

Table~\ref{tab:component_selection_counts} summarizes the numbers of retained PCA components in the three antennas and the numbers satisfying the selection criteria listed in Table \ref{tab:selection_parameters}. In total, 235,964 components are retained from the full data set, but only 81 are classified as non-shared after the cross-antenna comparison. After the harmonic, frequency-domain, and SNR criteria are applied jointly, no final candidate remains. We also visually inspect the all the PCA summary plots with top components to confirm whether periodicities actually appear in these components with no frequency comb structure. Figure \ref{fig:example_PCA_summary} illustrates such an example PCA summary plot. No significant periodic signal appears in the components.

\begin{deluxetable*}{lcccc}
\tablecaption{Numbers of retained PCA components and components satisfying each selection criterion for the three Chang'e-4 antennas.\label{tab:component_selection_counts}}
\tablehead{
\colhead{Condition} & \colhead{A} & \colhead{B} & \colhead{C} & \colhead{Total}
}
\startdata
Retained components     & 81605 & 72388 & 81971 & 235964 \\
Harmonic criterion      & 9116  & 8240  & 9029  & 26385  \\
Period-SNR criterion    & 43903 & 39030 & 44252 & 127185 \\
Non-shared components   & 0     & 41    & 40    & 81     \\
FFT criterion           & 689   & 733   & 542   & 1964   \\
ACF criterion           & 5132  & 5117  & 6649  & 16898  \\
Final candidates        & 0     & 0     & 0     & 0      \\
\enddata
\end{deluxetable*}

\begin{figure}[htpb]
  \centering
  \includegraphics[width=\textwidth]{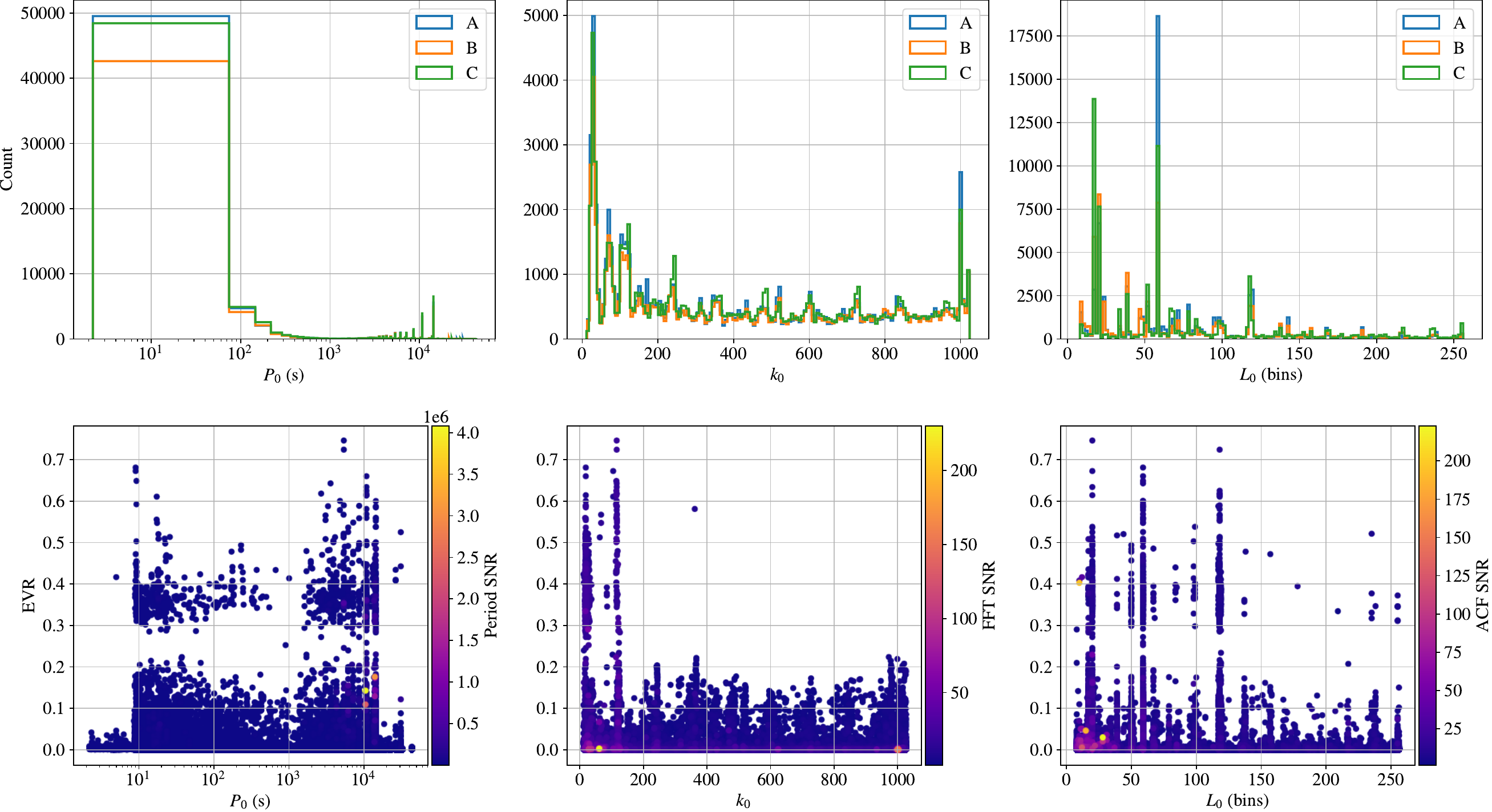}
  \caption{Statistical distributions of the component periodicity and spectral behaviors. The top row shows the dominant period $P_0$ (left), the frequency repetition scale $k_0$ (middle), and the ACF peak lag $L_0$ in frequency bins (right) for the components in antennas A, B, and C. The bottom row shows the corresponding EVR-$P_0$, EVR-$k_0$, and EVR-$L_0$ relations, with the color indicating the corresponding SNR.}
  \label{fig:component_distrubution}
\end{figure}

We analyze the statistical distribution of periodicity and spectral behaviors for the components as well, which is shown in Figure \ref{fig:component_distrubution}. In all three antennas, these quantities are not uniformly distributed, but instead cluster around several preferred ranges, indicating that both periodic and repeated frequency-domain structures appear only at certain scales. The concentration toward short periods may reflect not only a larger number of short-period components, but also the contribution of harmonic peaks, which shift part of the periodogram power from a longer fundamental period to shorter periods at $P_0/n$. The distributions of frequency-domain structures diagnostics show peaks at $k_0 \sim 28$ and $L_0 \sim 17$ bins, which is consistent with the comb structure found in \cite{2023RaSc...5807595L} at low-frequency band. The EVR relations further show that most of these distributions are associated with low-EVR components. Those components with high period-SNR are long-period components, and the components with high FFT-SNR/ACF-SNR are distributed at short $k_0$ and $L_0$ with small EVR, indicating these high-SNR components are probably contributed by comb-like RFI.

\section{Discussion} \label{sec:Discussion}
\subsection{Sensitivity}
The system equivalent flux density (SEFD) for a radio antenna can be written as 
\begin{equation}
  \mathrm{SEFD}=\frac{2k_\mathrm{B}T_\mathrm{sys}}{A_\mathrm{eff}},
  \label{SEFD}
\end{equation}
where $k_\mathrm{B}$ is the Boltzmann constant, $T_\mathrm{sys}$ is the system temperature, and $A_\mathrm{eff}$ is the effective collecting area. For an ideal dipole, $A_\mathrm{eff}=3\lambda^2/8\pi$. The minimum detectable flux density $S_{\min}$ can be given by
\begin{equation}
  S_{\min}=\mathrm{SNR}_{\min}\frac{\mathrm{SEFD}}{\sqrt{\delta \nu_{\rm ch} \tau_\mathrm{eff}}},
  \label{NBS_min}
\end{equation}
where $\mathrm{SNR}_{\min}$ is the SNR threshold, $\delta \nu_{\rm ch}$ is the frequency channel resolution, $\tau_\mathrm{eff}=D\tau_\mathrm{obs}$ is the effective observing time, with $\tau_\mathrm{obs}$ being the observation duration. In this work, the SNR threshold we adopt is 10, and the frequency channel resolution is about 19 kHz. For a typical observation duration of 12 hr, with typical duty cycle 0.04 in the search, we calculate $S_{\min}\sim 3\times10^{-13} \, \mathrm{W \cdot m^{-2} \cdot Hz^{-1}}$. We can also calculate the minimum detectable equivalent isotropic radiated power (${\rm EIRP_{\min}}$) by 
\begin{equation}
\mathrm{EIRP}_{\min} \approx 4\pi d^2 S_{\min} \delta\nu_{\rm ch}
\end{equation}
For a source of ETI signal candidates at $d=10$ pc, we calculate $\mathrm{EIRP}_{\min} \approx 1.95\times10^{28}$ W.
\subsection{PCA Characterization in the Analysis}

\begin{figure}[htpb]
  \centering
  \includegraphics[width=0.75\textwidth]{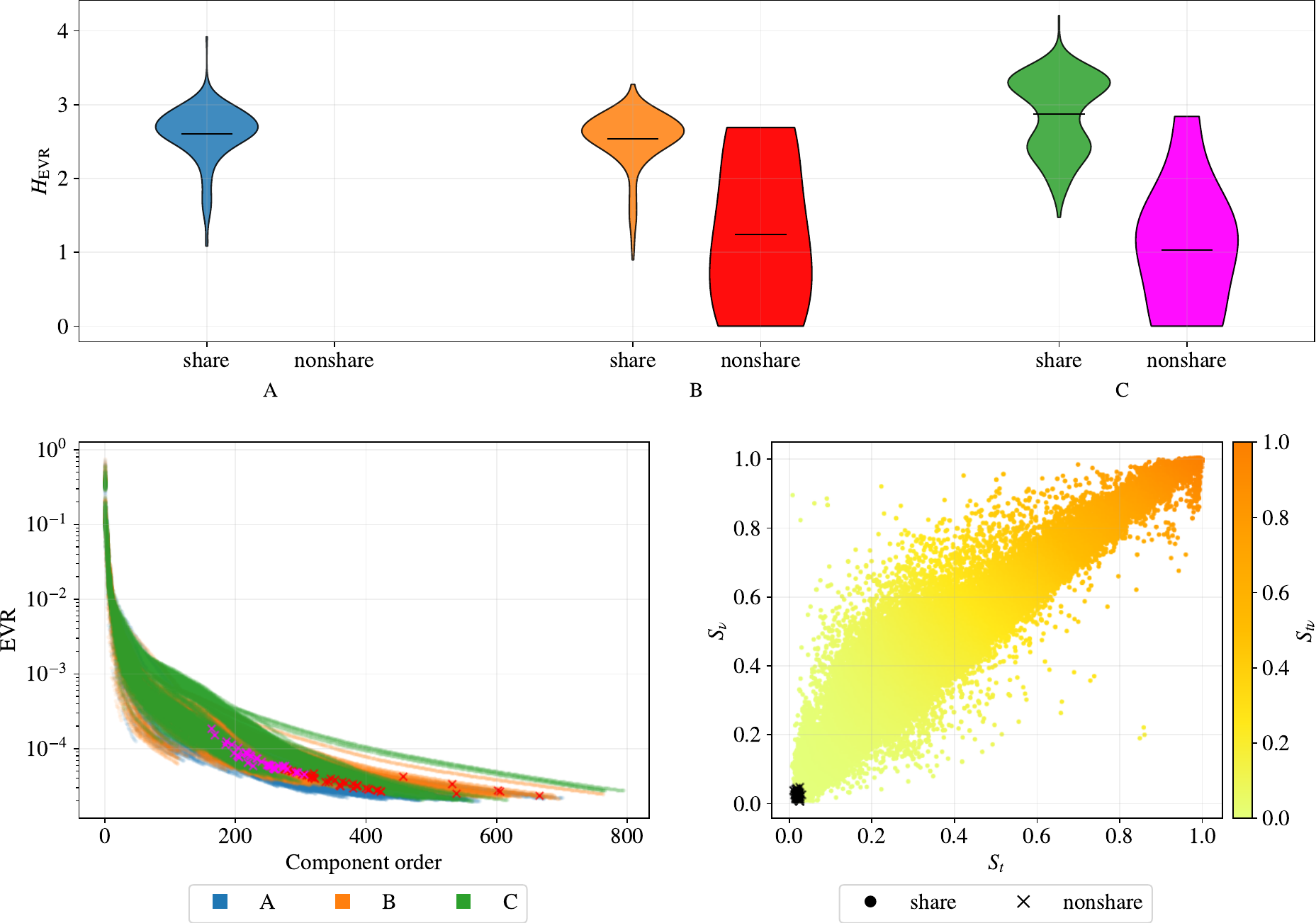}
  \caption{Statistical characterization of PCA components and their cross-antenna properties. Top: violin plot of the EVR entropy $H_{\rm EVR}$ for antennas A, B, and C. Bottom left: EVR versus component order scatters for all retained components. Bottom right: distribution of the time and frequency basis similarities, colored by $S_{t\nu}=S_tS_f$. The shared and non-shared components are plotted with dot and cross, respectively.}
  \label{fig:PCA_properties}
\end{figure}

To further characterize the role of the PCA decomposition in our analysis, we examine how the variance is distributed among components and how the resulting components are separated as shared and non-shared ones in the basis alignment. This variance concentration can be quantified by the EVR entropy, 
\begin{equation}
  H_{\rm EVR}=-\sum_i p_i\ln p_i,\qquad p_i=\frac{{\rm EVR}_i}{\sum_j {\rm EVR}_j}.
\end{equation}
Lower $H_{\rm EVR}$ values indicate that the variance is concentrated in a smaller number of top components, corresponding to a more concentrated variance distribution, whereas higher $H_{\rm EVR}$ values indicate that the variance is spread among a larger number of components. The $H_{\rm EVR}$ distributions show a broadly similar overall trend among antennas A, B, and C (See top row of Figure \ref{fig:PCA_properties}). Most observations cluster between moderate to relatively low entropy levels, while high $H_{\rm EVR}$ cases are rare. Specifically, antenna C tends to exhibit slightly larger $H_{\rm EVR}$ values and a broader distribution, suggesting that its variance is less concentrated than in A and B. The non-shared subsets of antenna B and C are distributed toward lower $H_{\rm EVR}$, while this trend should be interpreted cautiously because of the much smaller sample size.

The EVR distributions in component rank orders, illustrated in the bottom left panel of Figure \ref{fig:PCA_properties}, reflect that the variance drops rapidly with component rank order in all antennas. Importantly, the non-shared components are largely confined to the low-EVR with high rank orders, suggesting that potentially non-shared structures do not dominate the total variance. The cross-antenna alignment distributions in bottom right panel of Figure \ref{fig:PCA_properties} show a clear positive trend, indicating that components with stronger similarity in the time basis also tend to exhibit stronger similarity in the frequency basis. At the same time, the non-shared subset is concentrated in the lower-left corner, implying that these components are typically non-shared in both bases rather than in only one of them.

\section{Conclusion} \label{sec:Conclusion}

We conducted a channel-width periodic radio signal search using the CE-4 lunar-farside low-frequency radio observation data. This is the first SETI study based on radio observations obtained from the lunar farside. We introduce PCA to characterize potential periodic structures in the CE4 dynamic spectra. Using a component-level framework based on PCA decomposition, cross-antenna basis alignment, and periodogram/FT/ACF diagnostics, we examined the temporal periodicity and frequency-domain repetition behavior of the retained components, to determine potential non-shared periodic signals. After all selection criteria applied to each component in the search, no candidate periodic signal was found in this search. Our results suggest that the dominant structured features in the current data are more likely associated with shared instrumental RFIs.

The lunar farside provides one of the most favorable environments for radio observations, and upcoming lunar exploration programs are likely to make such observations increasingly systematic. We look forward to the LFRS 
of the CE4 mission launching a new round of observations to obtain more data for SETI research. Some SETI related projects has already pointed to planned radio telescopes on the lunar farside. The planning China's ``Guanghanjing'' Lunar Farside SETI Telescope (GLFST) is expected to be equipped on the Chang'E-8 lander, with frequency range of 0.1 MHz-2.7 GHz, which generally covers the HF, VHF, UHF and L band. The Lunar Surface Electromagnetics Experiment (LuSEE-Night) project\footnote{\url{https://www.cosmo.bnl.gov/node/5}}, under development by NASA and UC Berkeley, attempts to measure the eraly age of the Universe and carry out SETI observation. \cite{2021RSPTA.37990566C} also propose a lunar orbit array for discovering sky at the longest wavelength (DSL) project, which can be also applied in SETI research. The proposed Lunar Farside Technosignature and Transient Telescope (LFT3) is designed for lunar-farside observations for radio emissions from known and unknown sources \citep{2025usnc.conf...46D}. In parallel, Future Chinese lunar mission plans, including Chang'E-8\footnote{\url{https://www.cnsa.gov.cn/english/n6465652/n6465653/c10670293/content.html}} and the broader international lunar research station \citep{WU2023101537} roadmap, already point to planned radio astronomy payloads. NASA has started deploying low-frequency lunar radio spectrometers such as the Lunar Surface of the photoElectron Sheath (ROLSES) through the CLPS framework \citep{2025pds..data...43G}. These developments are consistent with the view that lunar-based radio observations can provide a particularly favorable environment for SETI, and also suggest that the Moon, and especially the lunar farside, may become an increasingly important platform for future SETI studies.

\begin{acknowledgments}
We thank Yan Su for the kind and useful discussions and suggestions. We thank the Chang'E-4 payload team for mission operations and China National Space Administration for providing the Chang'E-4 data that made this study possible. This work was supported by National Key R\&D Program of China, No.2024YFA1611804 and the China Manned Space Program with grant No. CMS-CSST-2025-A01. The Chang'E-4 data used in this work is processed and produced by ``Ground Research and Application System (GRAS) of China's Lunar and Planetary Exploration Program,  provided by China National Space Administration (\url{https://moon.bao.ac.cn})''. 
\end{acknowledgments}

%

\software{Python, NumPy \citep{2020Natur.585..357H}, Pandas \citep{mckinney2010,the_pandas_development_team_2022_7344967}, Matplotlib \citep{2007CSE.....9...90H}, SciPy \citep{2020SciPy-NMeth}, scikit-learn \citep{2011JMLR...12.2825P}}


\appendix

\section{Theory of Cross-antenna Basis Alignment}\label{Appendix:basis_alignment}

For two antennas labeled $a$ and $b$, let
$U_a \in \mathbb{R}^{N_t \times N_a}$ and $V_a \in \mathbb{R}^{N_\nu \times N_a}$
denote the retained temporal and frequency bases of antenna $a$, and similarly
$U_b \in \mathbb{R}^{N_t \times N_b}$ and $V_b \in \mathbb{R}^{N_\nu \times N_b}$
for antenna $b$.
Because the retained component numbers need not be identical between antennas,
the alignment is performed on the common retained dimension.

We first characterize the overall alignment between the retained PCA spaces of two antennas to provide a global measure of whether the dominant modes in two antennas are broadly aligned. We form the overlap matrices
\begin{equation}
 Q_t = U_a^{\rm T} U_b,
\qquad
Q_\nu = V_a^{\rm T} V_b, 
\end{equation}
for the temporal and frequency bases, respectively, to quantify the alignment between the retained subspaces. The overall alignment of two retained subspaces is then summarized by the normalized squared Frobenius norm,
\begin{equation}
  S_{\rm sub} = \frac{\|Q\|_F^2}{N},
\end{equation}
where $Q$ denotes either $Q_t$ or $Q_\nu$, and $N$ is the dimension used in the alignment. If the two subspaces are well aligned, the corresponding overlaps are large. Then we carry out the cross-antenna component alignment. For the normalized temporal and frequency basis vectors pairs consisting of the $i$-th and $j$-th retained component $u_{a,i}, v_{a,i}$ and $u_{b,j}, v_{b,j}$, the temporal and frequency alignment measures of a matched pair can be written as 
\begin{equation}
  S_t(i,j) = \left|u_{a,i}^{\rm T} u_{b,j}\right|, \quad S_\nu(i,j) = \left|v_{a,i}^{\rm T} v_{b,j}\right|.
\end{equation}
To favor components that are aligned in both time and frequency, we define the joint alignment measure as
\begin{equation}
  S_{t\nu}(i,j) = S_t(i,j)S_\nu(i,j).
\end{equation}
For each component $i$ in antenna $a$, we search over all components $j$ in antenna $b$ and identify the best-aligned counterpart as the one that maximizes $S_{t\nu}(i,j)$. 

Under the null hypothesis that two normalized basis vectors are randomly oriented in a $d$-dimensional space, their inner product $x \in \mathbb{R}^d$ has zero mean and variance $1/d$, with density
\begin{equation}
  p(x)\propto (1-x^2)^{(d-3)/2}, \qquad -1\le x\le 1.
\end{equation}
Equivalently,
\begin{equation}
  x^2 \sim \mathrm{Beta}\left(\frac{1}{2},\frac{d-1}{2}\right).
\end{equation}
In the high-dimensional regime, the distribution of $x$ is well approximated by
\begin{equation}
  x \sim N(0,1/d).
\end{equation}
Accordingly, the rescaled squared inner product is approximately distributed as a chi-square variable with one degree of freedom,
\begin{equation}
  (d-1)x^2 \approx \chi^2_1.
\end{equation}
Applying this approximation to the temporal and frequency alignment measures, the temporal and frequency test statistics
\begin{equation}
  T_t = (N_t-1) S_t^2, \quad T_\nu = (N_\nu-1) S_\nu^2
\end{equation}
also satisfy approximate distribution $\chi^2_1$, and the same can be applied to the joint alignment statistic $T_{t\nu} = T_t + T_\nu$ as well. A small p-value indicates that the observed joint alignment in time and frequency is unlikely to arise from random orientation alone, and therefore supports the interpretation that the component is shared across antennas. Conversely, if no statistically significant aligned counterpart is found, the component is classified as non-shared.

The cross-antenna basis alignment described above is not intended to establish strict physical identity between components. Rather, it provides a quantitative way to determine whether similar temporal and spectral modes recur across antennas. Those components that remain non-shared after alignment are of greater interest for the identification of non-RFI periodic signals.


\bibliography{sample701}{}
\bibliographystyle{aasjournalv7}



\end{document}